\documentclass[twocolumn,showpacs,showkeys,reprint,amsmath,amssymb,aps]{revtex4-2}

\usepackage{graphicx}
\usepackage{dcolumn}
\usepackage{bm}
\usepackage{tabularx}
\usepackage{amsmath}
\usepackage{xcolor}
\usepackage{float}

\emergencystretch=\maxdimen
\begin{document}
	
\title{Ultralow-temperature heat transport evidence for residual density of states in the superconducting state of CsV$_3$Sb$_5$}
	
\author{C. C. Zhao,$^{1,2}$ L. S. Wang,$^{1}$ W. Xia,$^{3,4}$ Q. W. Yin,$^{5}$ H. B. Deng,$^{6}$ G. W. Liu,$^{6}$ J. J. Liu,$^{7}$ X. Zhang,$^1$ J. M. Ni,$^1$ Y. Y. Huang,$^1$ C. P. Tu,$^1$ Z. C. Tao,$^{3}$ Z. J. Tu,$^{5}$ C. S. Gong,$^{5}$ Z. W. Wang,$^{7}$ H. C. Lei,$^{5,\dag}$ Y. F. Guo, $^{3,\ddag}$ X. F. Yang,$^{1,\sharp}$ J. X. Yin,$^{6,\S}$ and S. Y. Li$^{1,2,8,9,*}$}
	
\affiliation
{$^1$State Key Laboratory of Surface Physics, Department of Physics, Fudan University, Shanghai 200438, China\\
$^2$Shanghai Research Center for Quantum Sciences, Shanghai 201315, China\\
$^3$School of Physical Science and Technology, ShanghaiTech University, Shanghai 201210, China\\
$^4$ShanghaiTech Laboratory for Topological Physics, Shanghai 201210, China\\
$^5$Department of Physics and Beijing Key Laboratory of Opto-electronic Functional Materials and Micro-nano Devices, Renmin University of China, Beijing 100872, China\\
$^6$Department of Physics, Southern University of Science and Technology, Shenzhen 518055, China\\
$^7$Centre for Quantum Physics, Key Laboratory of Advanced Optoelectronic Quantum Architecture and Measurement (MOE), School of Physics, Beijing Institute of Technology, Beijing 100081, China\\
$^8$Shanghai Branch, Hefei National Laboratory, Shanghai 201315, China\\
$^9$Collaborative Innovation Center of Advanced Microstructures, Nanjing 210093, China
}
	
\date{\today}
	
\begin{abstract}
The V-based kagome superconductors $A$V$_3$Sb$_5$ ($A$ = K, Rb, and Cs) host charge density wave (CDW) and a topological nontrivial band structure, thereby provide a great platform to study the interplay of superconductivity (SC), CDW, frustration, and topology. Here, we report ultralow-temperature thermal conductivity measurements on CsV$_3$Sb$_5$ and Ta-doped Cs(V$_{0.86}$Ta$_{0.14}$)$_3$Sb$_5$ and scanning tunneling microscopy (STM) measurements on CsV$_3$Sb$_5$. The finite residual linear term of thermal conductivity at zero magnetic field suggests the existence of a residual density of states (DOS) in the superconducting state of CsV$_3$Sb$_5$. This is supported by the observation of non-zero conductance at zero bias in STM spectrum at an electronic temperature of 90 mK. However, in Cs(V$_{0.86}$Ta$_{0.14}$)$_3$Sb$_5$, which does not have CDW order, there is no evidence for residual DOS. These results show the importance of CDW order for the residual DOS, and a nodal $s$-wave gap or residual Fermi arc may be the origin of the residual DOS in such an unusual multiband kagome superconductor, CsV$_3$Sb$_5$.
		
\end{abstract}
	
\maketitle

Superconductivity (SC) has been discovered in a family of vanadium-based kagome compounds, $A$V$_3$Sb$_5$ ($A$ = K, Rb, and Cs) \cite{PRM-Vcompounds,Cs-Z2-SC,K-SC,Rb-SC}. The superconducting transition temperature $T_c$ are 0.93, 0.92, and 3 K for $A$ = K, Rb, and Cs, respectively \cite{K-SC,Rb-SC,Chen-PRX}. Although there are no local magnetic moments \cite{Nonmagnetic}, these V-based superconductors exhibit charge density wave (CDW) order at 78, 103, and 94 K \cite{PRM-Vcompounds,Cs-Z2-SC,K-SC,Rb-SC}. Various measurements have revealed a 2 $\times$ 2 in-plane superlattice modulation associated with the CDW transition in $A$V$_3$Sb$_5$, however, it is still controversial whether the three-dimensional CDW is 2 $\times$ 2 $\times$ 2 or 2 $\times$ 2 $\times$ 4 along the $c$ axis \cite{nm-stm-KVS,PRX-CDW-ortiz,nature-CDW-CVS,nature-STM,Chen-PRX,PRB-CDW-AVS,PRX-CDW,nc-CDW}. Within the CDW state, many exotic properties, such as giant anomalous Hall effect \cite{nm-stm-KVS,AHE,PRB-AHE,arxiv-AHE}, the time-reversal symmetry breaking \cite{nature-TRSB,arxiv-Kerr,nm-Yin}, possible loop current ordering \cite{SciBull-Hu,np-nematic} and electronic nematicity \cite{np-nematic,nature-nematic-Chen,nc-twofold}, have been discovered.

When $A$V$_3$Sb$_5$ enters into a superconducting state, its properties also attract considerable attention. Two superconducting domes can be obtained with increasing pressure or doping \cite{Zhao-arxiv-08356,Zhu-PRB-pressure,Cheng-PRL-pressure,nc-CVTS-nematic,scibull-CVTS,PRM-Sb-Sn}, and the CDW order was intertwined with the superconductivity in an unconventional manner \cite{song-nc,NSR-AVS-Hu,nature-Chen-pressure}. The presence of an unusual pair density wave (PDW) makes the complexity of a superconducting state \cite{nature-STM,PRB-PDW,nature-PDW,PDW-RVS-Yin,arxiv-2408.02890-Yin}. Regarding the superconducting gap structure, some early studies claimed that CsV$_3$Sb$_5$ is fully gapped \cite{Yuan-ACPMA,Lei-miuSR-npj,nodeless-communphys,nc-bulk-conventional,NMR-Lei}, whereas others have suggested nodal superconductivity in $A$V$_3$Sb$_5$ \cite{Zhao-arxiv-08356,nc-miuSR-node}. The latest angle-resolved photoemission spectroscopy (ARPES) measurements of high-quality CsV$_3$Sb$_5$ single crystals at 2 K revealed that the SC gap in the $\beta$ Fermi surface (FS) of CsV$_3$Sb$_5$ is highly anisotropic, and the minimum SC gap is considerably smaller than the experimental energy resolution, suggesting the possible presence of a node \cite{arxiv2404}. However, some scanning tunneling microscopy (STM) studies observed a non-zero conductance at zero bias \cite{Feng-PRL-s,Chen-PRX,nature-STM}, which may be related to the residual Fermi arc \cite{nature-PDW}. Such complexity of the superconducting state does not appear in the Ta-doped CsV$_3$Sb$_5$ with a higher $T_c$ of 5 K and without a CDW \cite{nm-Yin,nc-CVTS,nature-CVTS-nodeless}.  For example, the ARPES measurements on Cs(V$_{0.86}$Ta$_{0.14}$)$_3$Sb$_5$ concluded that the superconducting gaps are isotropic and lack nodes \cite{nature-CVTS-nodeless}. Similarly, a U-shaped superconducting gap characterized by a flat bottom and zero density of states (DOS) at zero bias was observed in the STM spectrum of CsV$_{2.6}$Ta$_{0.4}$Sb$_5$ \cite{nm-Yin,nc-CVTS}.

In this study, we conduct ultralow-temperature thermal conductivity and STM measurements of CsV$_3$Sb$_5$ to investigate its superconducting gap structure. The thermal conductivity in zero magnetic field shows a clear residual linear term $\kappa_0/T$, indicating the existence of a residual DOS within the superconducting state of CsV$_3$Sb$_5$. This is supported by the STM result, which provides a significant non-zero DOS at zero bias with an electronic temperature of 90 mK and a lattice temperature of 30 mK. Moreover we also measure the ultralow-temperature thermal conductivity of its substituted counterpart, Cs(V$_{0.86}$Ta$_{0.14}$)$_3$Sb$_5$, which manifests no $\kappa_0/T$ in zero field and a exhibits completely different field dependence of $\kappa_0/T$ than that of CsV$_3$Sb$_5$. We discuss the origin of the residual DOS in CsV$_3$Sb$_5$ from the perspective of nodal gap and the interplay between CDW and superconductivity.
	
\begin{figure*}
\centering
\includegraphics[clip,width=13.5cm]{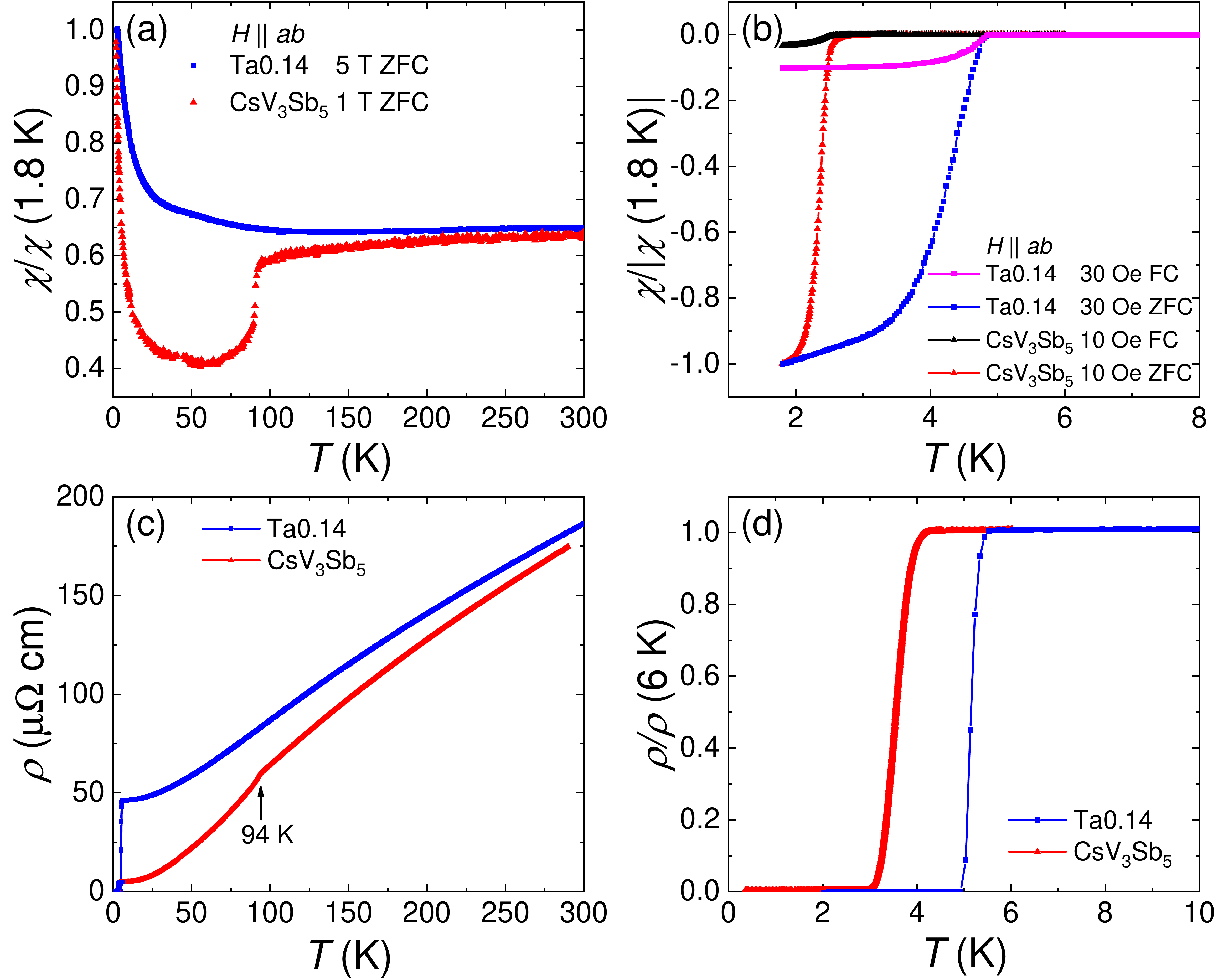}
\caption{(a) Temperature dependence of the normalized dc magnetization for CsV$_3$Sb$_5$ and Cs(V$_{0.86}$Ta$_{0.14}$)$_3$Sb$_5$ single crystals. (b) Normalized dc magnetization for CsV$_3$Sb$_5$ and Cs(V$_{0.86}$Ta$_{0.14}$)$_3$Sb$_5$ at low temperatures, with zero-field and field cooling modes. (c) Temperature dependence of in-plane resistivity for CsV$_3$Sb$_5$ and Cs(V$_{0.86}$Ta$_{0.14}$)$_3$Sb$_5$ single crystals. The arrow denotes the charge density wave order at 94 K for CsV$_3$Sb$_5$. (d) Normalized resistivity curves at low temperatures showing clear superconducting transitions.}
\end{figure*}

\begin{figure}
\includegraphics[clip,width=6.0cm]{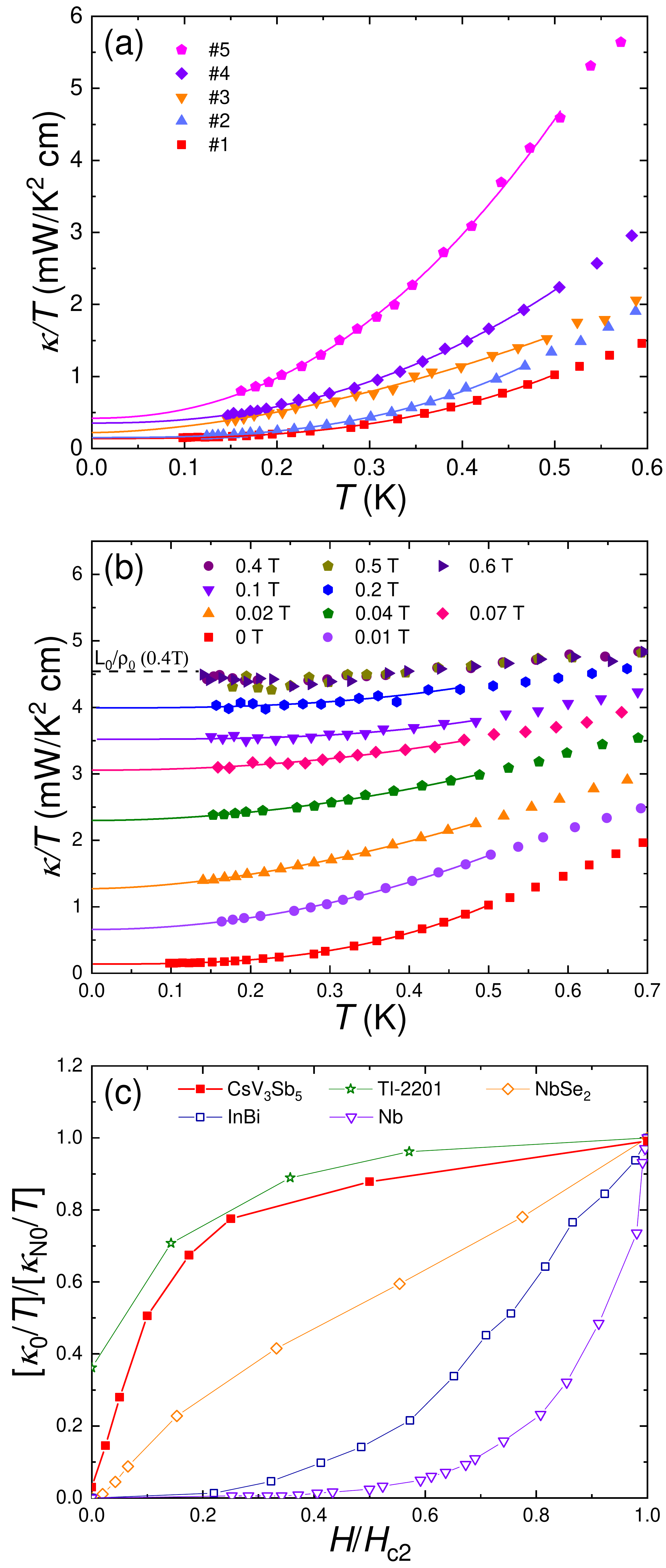}
\caption{(a) Temperature dependence of the in-plane thermal conductivity for five CsV$_3$Sb$_5$ single crystals in zero field. The solid line represents a fit to $\kappa/T = a+bT^{\alpha - 1}$ below 0.5 K. (b) The thermal conductivity of CsV$_3$Sb$_5$ sample \#1 in magnetic fields up to 0.6 T. The dashed line is the normal-state Wiedemann-Franz law expectation $L_0/\rho_0$(0.4 T), with the Lorenz number $L_0 = 2.45 \times 10^{-8}$ W $\Omega$ K$^{-2}$ and $\rho_0$(0.4 T) = 5.40 $\mu \Omega$ cm. (c) Normalized residual linear term $\kappa_0/T$ of CsV$_3$Sb$_5$ as a function of $H/H_{\rm_{c2}}$, with bulk $\mu_0H{\rm_{c2}}$ = 0.4 T. Similar data of the clean $s$-wave superconductor Nb \cite{Nbkappa}, the dirty $s$-wave superconducting alloy InBi \cite{InBikappa}, the multiband $s$-wave superconductor NbSe$_2$ \cite{Boaknin2003}, and an overdoped $d$-wave cuprate superconductor Tl-2201 \cite{Proust2002} are shown for comparison.}
\end{figure}

Figure 1(a) shows the normalized DC magnetization of the CsV$_3$Sb$_5$ and Cs(V$_{0.86}$Ta$_{0.14}$)$_3$Sb$_5$ single crystals measured at 1 T and 5 T in zero-field cooling modes, respectively. The sharp drop at 94 K in CsV$_3$Sb$_5$ corresponds to a previously reported CDW transition \cite{Cs-Z2-SC}, whereas the CDW was suppressed by Ta doping in Cs(V$_{0.86}$Ta$_{0.14}$)$_3$Sb$_5$ \cite{nm-Yin}. In Fig. 1(c), the CDW transition manifests as a resistivity kink at 94 K for CsV$_3$Sb$_5$, and no anomaly is observed in the resistivity of Cs(V$_{0.86}$Ta$_{0.14}$)$_3$Sb$_5$. For the low-temperature magnetization results shown in Fig. 1(b), without considering the demagnetization factor, the calculated superconducting volume fractions are 128\% and 106\% for CsV$_3$Sb$_5$ and Cs(V$_{0.86}$Ta$_{0.14}$)$_3$Sb$_5$, respectively. These values demonstrate the bulk superconductivity of the samples. At low temperature, the resistivity of both CsV$_3$Sb$_5$ and Cs(V$_{0.86}$Ta$_{0.14}$)$_3$Sb$_5$ show a superconducting transition at 3.0 K and 4.9 K (defined at zero resistivity), respectively, as shown in Fig. 1(d). These $T_c$ values obtained from the resistivity are slightly higher than those obtained from magnetization in Fig. 1(b). Further details regarding the sample growth, low temperature resistivity in field, upper critical field, and measurement procedure can be found in the Supplemental Material \cite{SM-CVS}.

Ultralow-temperature heat transport measurement is a well-established bulk technique for probe superconducting gap structures \cite{Shakeripour2009}. Figure 2(a) shows the in-plane thermal conductivity of five CsV$_3$Sb$_5$ single crystals in zero field. At ultralow temperature, the thermal conductivity can be fitted to $\kappa/T$ = $a$ + $bT^{\alpha-1}$ \cite{Sutherland2003,SYLi2008}, where the terms $aT$ and $bT^\alpha$ represent contributions from electrons and phonons, respectively. The power $\alpha$ is typically between 2 and 3, because of the specular reflections of phonons at the boundary \cite{Sutherland2003,SYLi2008}. For samples \#1-\#5 below 0.5 K, the fitting gives a finite residual linear term $(\kappa_0/T)_1 =$ 0.14 $\pm$ 0.002 mW K$^{-2}$ cm$^{-1}$, $(\kappa_0/T)_2 =$ 0.15 $\pm$ 0.004 mW K$^{-2}$ cm$^{-1}$, $(\kappa_0/T)_3 =$ 0.22 $\pm$ 0.038 mW K$^{-2}$ cm$^{-1}$, $(\kappa_0/T)_4 =$ 0.35 $\pm$ 0.013 mW K$^{-2}$ cm$^{-1}$, and $(\kappa_0/T)_5 =$ 0.42 $\pm$ 0.057 mW K$^{-2}$ cm$^{-1}$, respectively. Furthermore, it can be observed that most of the obtained power $\alpha$ shows a value larger than 3, with $\alpha_1$ = 3.87 $\pm$ 0.03, $\alpha_2$ = 3.85 $\pm$ 0.05, $\alpha_3$ = 2.70 $\pm$ 0.12, $\alpha_4$ = 3.28 $\pm$ 0.05, and $\alpha_5$ = 3.19 $\pm$ 0.08, respectively. This anomalously large $\alpha$ suggests the existence of thermally excited electrons with increasing temperature, even at such low temperatures. Considering that the $T_c$ of CsV$_3$Sb$_5$ is quiet low and the observation of a highly anisotropic superconducting gap in one of the complex FSs, a considerably small superconducting gap is gradually closed by the temperature. Although the second term $bT^{\alpha-1}$ may contain the contribution from thermally excited electrons, the finite residual linear term extrapolated from the ultralow-temperature thermal conductivity provides evidence for the existence of a residual DOS in the superconducting state of CsV$_3$Sb$_5$.	
	
For $s$-wave nodeless superconductors, there are no fermionic quasiparticles to conduct heat as $T \rightarrow 0$ because the FS is entirely gapped \cite{Sutherland2003,SYLi2008}. Therefore there is no residual linear term, $\kappa_0/T$, as observed for InBi and NbSe$_2$ \cite{InBikappa,Boaknin2003}. However, nodal superconductors exhibit a significant $\kappa_0/T$ at zero field owing to nodal quasiparticles \cite{Shakeripour2009}. For example, $\kappa_0/T$ of the overdoped ($T${$\rm_c$} = 15 K) $d$-wave cuprate superconductor Tl$_2$Ba$_2$CuO$_{6+\delta}$ (Tl-2201) is 1.41 mW K$^{-2}$ cm$^{-1}$, which is about $36\%$ of the normal-state value $\kappa{\rm_{N0}}/T$ \cite{Proust2002}. For clean Fe-based superconductor KFe$_2$As$_2$, $\kappa_0/T$ = 3.8 mW K$^{-2}$ cm$^{-1}$ is about $3.2\%$ of the normal-state value $\kappa{\rm_{N0}}/T$ \cite{KFe2As2}. For ultraclean YBa$_2$Cu$_3$O$_7$ (i.e., considerably low $\rho_0$ and significantly high $\kappa_0/T$ in the normal state), a universal $\kappa_0/T$ value 0.16 mW K$^{-2}$ cm$^{-1}$ was observed for $d$-wave superconducting gap \cite{YBCO}. Therefore, one possible origin of the residual DOS, indicated by the finite $\kappa_0/T$ of CsV$_3$Sb$_5$ in the zero field, may be the nodal quasiparticles. This issue will be discussed later.

Further information on the superconducting gap structure can be obtained by investigating the field-dependent behavior of $\kappa_0/T$ \cite{Shakeripour2009}. Figure 2(b) depicts the in-plane thermal conductivity of CsV$_3$Sb$_5$ sample \#1 under magnetic fields up to 0.6 T. The data were fitted using $\kappa/T = a+bT^{\alpha - 1}$ to determine the $\kappa_0/T$ for each magnetic field. The fitting of the data under $\mu_0H$ = 0.4 T yields $\kappa_0/T =$ 4.50 $\pm$ 0.02 mW K$^{-2}$ cm$^{-1}$. A further increasing field to 0.6 T does not increase the thermal conductivity, therefore, we assume 0.4 T as the bulk $H{\rm_{c2}}(0)$. This value is lower than that obtained from resistivity measurements. Furthermore, the value of $\kappa_0/T$ agrees well with the normal-state Wiedemann-Franz law expectation $L_0/\rho_0$(0.4 T) = 4.54 mW K$^{-2}$ cm$^{-1}$, with the Lorenz number $L_0 = 2.45 \times 10^{-8}$ W $\Omega$ K$^{-2}$ and $\rho_0$(0.4 T) = 5.40 $\mu \Omega$ cm (see the Supplemental Material \cite{SM-CVS} for more details). The validation of the Wiedemann-Franz law in the normal state confirmed the reliability of our thermal conductivity measurements.

\begin{figure}
\includegraphics[clip,width=6.0cm]{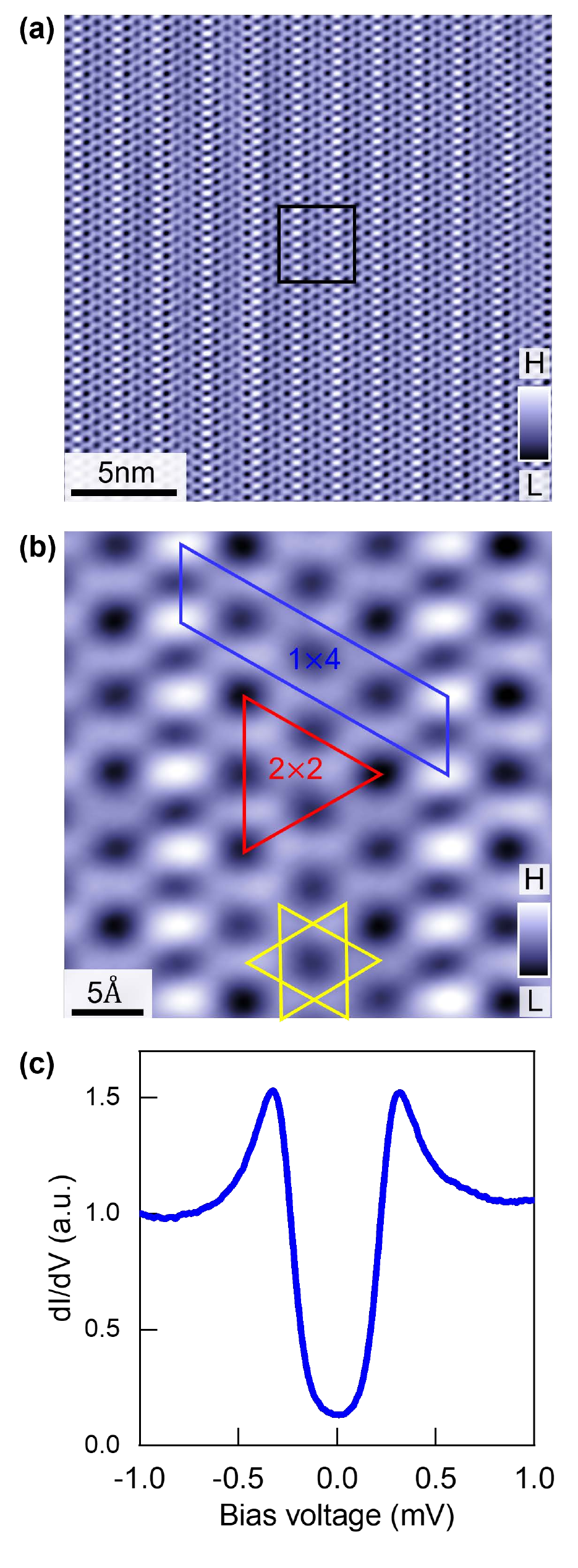}
\caption{(a) Typical STM images of the CsV$_3$Sb$_5$ surface. The scale bar is 5 nm. (b) Atomically resolved STM image of the sample surface taken in the dark box in (a). The blue quadrilateral and the red triangle represent the 1 $\times$ 4 and 2 $\times$ 2 supercells pattern, respectively. The yellow lines denote the underlying kagome lattice. (c) Spatially averaged $dI/dV$ spectrum of CsV$_3$Sb$_5$ measured at an electronic temperature of 90 mK.}
\end{figure}

The normalized values of $[\kappa_0/T]/[\kappa{\rm_{N0}}/T]$ as a function of $H/H{\rm_{c2}}$ for CsV$_3$Sb$_5$ is plotted in Fig. 2(c), with $\kappa{\rm_{N0}}/T$ = 4.54 mW K$^{-2}$ cm$^{-1}$ and $\mu_0H{\rm_{c2}}$ = 0.4 T. For comparison, similar data of the clean $s$-wave superconductor Nb \cite{Nbkappa}, the dirty $s$-wave superconducting alloy InBi \cite{InBikappa}, the multiband $s$-wave superconductor NbSe$_2$ \cite{Boaknin2003}, and an overdoped $d$-wave cuprate superconductor Tl-2201 \cite{Proust2002}, are also plotted. Note that the field dependence of $\kappa_0/T$ for CsV$_3$Sb$_5$ exhibits a rapid increase at low field. This is an indication of a nodal gap, as in Tl-2201, or of considerable small gaps in a multigap superconductor that can be suppressed by low fields.
	
To confirm the existence of residual DOS, we conducte ultralow-temperature STM measurements to investigate the SC state of CsV$_3$Sb$_5$. Figure 3(a) presents a typical STM image of the CsV$_3$Sb$_5$ surface, and Fig. 3(b) provides an enlarged view of the black box in Fig. 3(a). The atomically resolved lattice topographies shown in Fig. 3(b) further confirm the kagome lattice structure. At an electronic temperature of 90 mK, the coexistence of 1 $\times$ 4 and 2 $\times$ 2 supercells with the superconducting state was observed. However, for Cs(V$_{0.86}$Ta$_{0.14}$)$_3$Sb$_5$, the CDW is suppressed \cite{nm-Yin,nc-CVTS}. Furthermore, the superconducting gap of CsV$_3$Sb$_5$ exhibits a non-zero conductance at zero bias, as shown in Fig. 3(c). This appears to have a residual zero-energy DOS, which is consistent with our thermal conductivity measurements.

To determine whether the CDW order is essential for the residual DOS in CsV$_3$Sb$_5$, we measured the thermal conductivity of Cs(V$_{0.86}$Ta$_{0.14}$)$_3$Sb$_5$ in which the CDW is suppressed by Ta doping. The in-plane thermal conductivity of the Cs(V$_{0.86}$Ta$_{0.14}$)$_3$Sb$_5$ single crystal in zero field is shown in Fig. 4(a). The fitting of the data below 0.5 K with $\kappa/T = a+bT^{\alpha - 1}$ yields $\kappa_0/T$ = -0.4 $\pm$ 4 $\mu$W K$^{-2}$ cm$^{-1}$ and $\alpha$ = 2.28 $\pm$ 0.035. Considering our experimental error bar of $\pm$ 5 $\mu$W K$^{-2}$ cm$^{-1}$, the $\kappa_0/T$ of Cs(V$_{0.86}$Ta$_{0.14}$)$_3$Sb$_5$ in zero field is essentially zero. The power $\alpha$ also shows a normal value between two and three for phonons. Figure 4(b) shows the thermal conductivity of Cs(V$_{0.86}$Ta$_{0.14}$)$_3$Sb$_5$ under magnetic fields up to 2 T. Data below 0.5 K were fitted with $\kappa/T = a+bT^{\alpha - 1}$ to determine the $\kappa_0/T$ for each magnetic field. The fitting of the data under $\mu_0H$ = 1.8 T yields $\kappa_0/T =$ 0.51 $\pm$ 0.002 mW K$^{-2}$ cm$^{-1}$. Further increasing field to 2 T did not increase the thermal conductivity; therefore, we considered 1.8 T to be the bulk $H{\rm_{c2}}(0)$. This value of $\kappa_0/T$ agrees well with the expectation from the normal-state Wiedemann-Franz law $L_0/\rho_0$(1.8 T) = 0.53 mW K$^{-2}$ cm$^{-1}$, with the Lorenz number $L_0 = 2.45 \times 10^{-8}$ W $\Omega$ K$^{-2}$ and $\rho_0$(1.8 T) = 46.0 $\mu \Omega$ cm (see Supplemental Material \cite{SM-CVS} for more details). Figure 4(c) shows the normalized values of $[\kappa_0/T]/[\kappa{\rm_{N0}}/T]$ as a function of $H/H{\rm_{c2}}$ for Cs(V$_{0.86}$Ta$_{0.14}$)$_3$Sb$_5$. Here, we used $\kappa{\rm_{N0}}/T$ = 0.53 mW K$^{-2}$ cm$^{-1}$ and $\mu_0H{\rm_{c2}}$ = 1.8 T. The absence of $\kappa_0/T$ in zero field and the slow field dependence of $\kappa_0/T$ suggests nodeless superconductivity in Cs(V$_{0.86}$Ta$_{0.14}$)$_3$Sb$_5$, with no residual DOS. This is consistent with the STM results, which show a flat bottom with zero DOS \cite{nm-Yin,nc-CVTS}. Similar STM results were obtained for CsV$_{2.73}$Ti$_{0.27}$Sb$_5$ ($T_c$ = 3.7 K) \cite{scibull-CVTS}, showing that the effect of Ta doping is not unique in suppressing the CDW order and the residual DOS. Consequently, the CDW order in the pristine CsV$_3$Sb$_5$ seems to be essential for the existence of a residual DOS in zero field.

\begin{figure}
		
\includegraphics[clip,width=6.0cm]{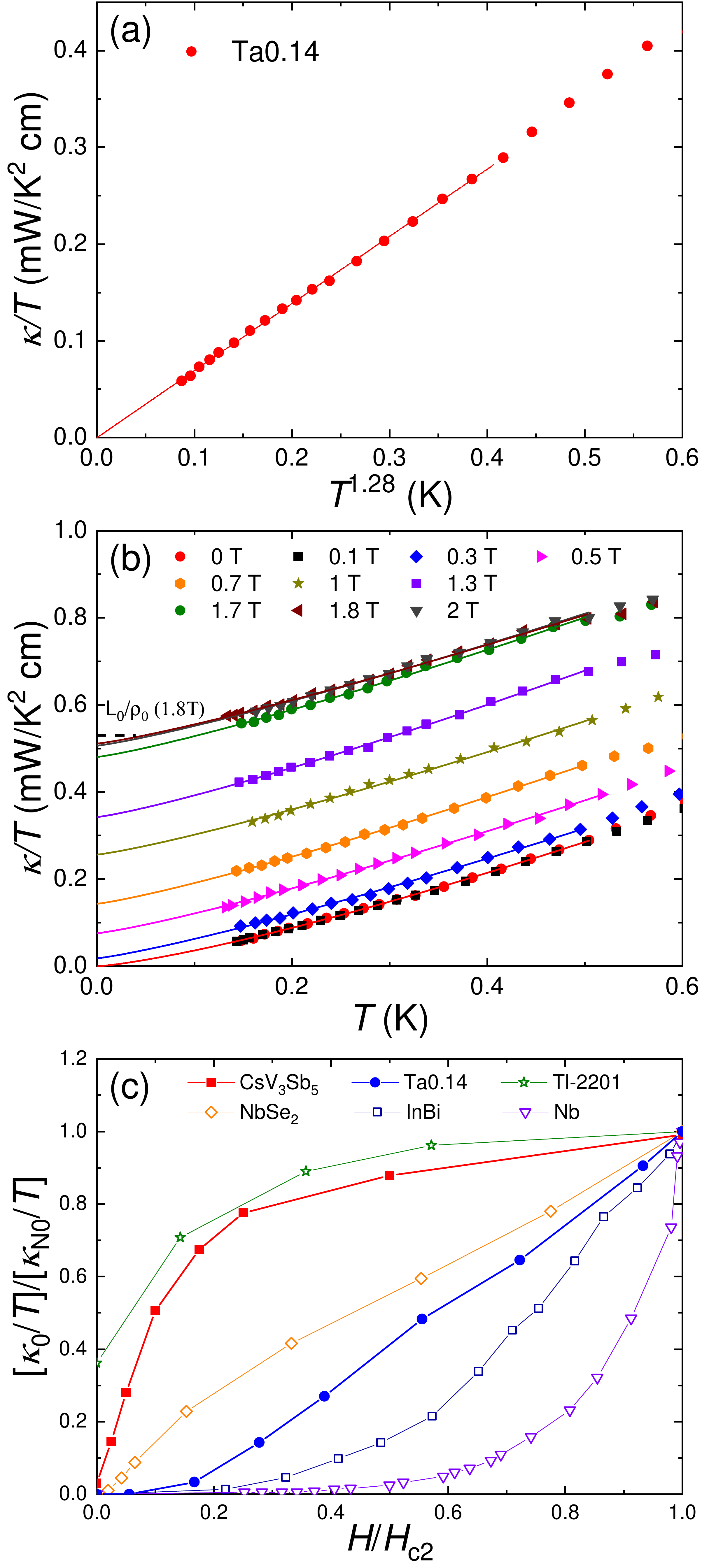}
\caption{(a) Temperature dependence of the in-plane thermal conductivity for Cs(V$_{0.86}$Ta$_{0.14}$)$_3$Sb$_5$ single crystal in zero field. The solid line represents a fit to $\kappa/T = a+bT^{\alpha - 1}$ below 0.5 K, which gives $\kappa_0/T =$ -0.4 $\pm$ 4 $\mu$W K$^{-2}$ cm$^{-1}$ and $\alpha$ = 2.28 $\pm$ 0.035. (b) The thermal conductivity of Cs(V$_{0.86}$Ta$_{0.14}$)$_3$Sb$_5$ in magnetic fields up to 2 T. The dashed line is the normal-state Wiedemann-Franz law expectation $L_0/\rho_0$(1.8 T), with the Lorenz number $L_0 = 2.45 \times 10^{-8}$ W $\Omega$ K$^{-2}$ and $\rho_0$(1.8 T) = 46.0 $\mu \Omega$ cm. (c) Normalized $\kappa_0/T$ of Cs(V$_{0.86}$Ta$_{0.14}$)$_3$Sb$_5$ as a function of $H/H_{\rm_{c2}}$, with bulk $H{\rm_{c2}}$ = 1.8 T. Similar data of the pristine CsV$_3$Sb$_5$, the clean $s$-wave superconductor Nb \cite{Nbkappa}, the dirty $s$-wave superconducting alloy InBi \cite{InBikappa}, the multiband $s$-wave superconductor NbSe$_2$ \cite{Boaknin2003}, and an overdoped $d$-wave cuprate superconductor Tl-2201 \cite{Proust2002} are shown for comparison.}
\end{figure}	
	
Subsequently, we discuss the origin of the residual DOS in CsV$_3$Sb$_5$. As aforementioned, such a residual DOS usually comes from nodal quasiparticles. However, several phase-sensitive experiments \cite{Feng-PRL-s,Yuan-ACPMA,Lei-miuSR-npj,NMR-Lei} indicate the absence of sign reversal of the SC order parameter, and low-temperature STM with high resolution suggested a flat residual in-gap state instead of a V-shape or full gap \cite{nature-PDW,PDW-RVS-Yin}. These observations caused an outstanding paradox that has not been encountered in other superconducting systems. The underlying reason for this may be that CsV$_3$Sb$_5$ is an unusual multiband superconductor. Recently, ARPES measurements on high-quality CsV$_3$Sb$_5$ single crystals at 2 K showed that the SC gaps in the $\alpha$ FS (from Sb 5$p$ orbital) and $\delta$ FS (from native V 3$d$ orbital) are isotropic, whereas the SC gap in the $\beta$ FS (from induced V 3$d$ orbital) is highly anisotropic \cite{arxiv2404}. It was interpreted that a nodal $s$-wave pairing state, which shows a node along $\Gamma$K direction and a maximum along V-V bond direction ($\Gamma$M), is highly consistent with the ARPES results \cite{arxiv2404}. Such a nodal $s$-wave gap, without sign reversal of SC order parameter, is consistent with our thermal conductivity results.
	
Another possible scenario is the strong anisotropic Cooper pairing of the tantalizing PDW order induced by the intertwining between a conventional wave SC and the CDW. The CDW occurs in the V $d$-orbital \cite{XJ Zhou-nc-cdw}, whereas the SC develops in the Sb $p$-orbital \cite{nature-PDW,PDW-RVS-Yin,arxiv2404}. The PDW state is a finite momentum-paired state with the momentum of the CDW or related vectors, and only when the electrons on the FS are connected by the PDW vector is the gap opened at the Fermi level. Electrons not connected by the PDW vector are left as Bogoliubov Fermi states, such as the recently reported residual Fermi arcs \cite{nature-PDW,PDW-RVS-Yin,arxiv-2408.02890-Yin}. Therefore, these residual Fermi arcs may be the origin of the residual DOS in CsV$_3$Sb$_5$ and contribute to the finite $\kappa_0/T$ in zero field.

In summary, we explore the superconducting gap structure of the V-based kagome superconductor CsV$_3$Sb$_5$ and its Ta-doped counterpart Cs(V$_{0.86}$Ta$_{0.14}$)$_3$Sb$_5$ through ultralow-temperature thermal conductivity measurements, together with the ultralow-temperature STM measurement of CsV$_3$Sb$_5$. The presence of a finite  $\kappa_0/T$ in a zero magnetic field and non-zero conductance at zero bias observed in STM demonstrate the existence of residual zero-energy DOS in CsV$_3$Sb$_5$. However, no residual DOS was observed for Cs(V$_{0.86}$Ta$_{0.14}$)$_3$Sb$_5$, indicating the importance of CDW order for the residual DOS. For this unusual multiband superconductor, a nodal $s$-wave gap or resudual Fermi arcs may be the origin of the residual DOS in CsV$_3$Sb$_5$.
	
This work was supported by the Natural Science Foundation of China (Grant No. 12174064), the Shanghai Municipal Science and Technology Major Project (Grant No. 2019SHZDZX01), and the Innovation Program for Quantum Science and Technology (Grant No. 2024ZD0300104). Y. F. Guo was supported by the Major Research Plan of the National Natural Science Foundation of China (No. 92065201) and the Program for Professor of Special Appointment (Shanghai Eastern Scholar). H. C. Lei was supported by National Natural Science Foundation of China (Grant No. 12274459), Ministry of Science and Technology of China (Grant Nos. 2022YFA1403800 and  2023YFA1406500).
	
C. C. Zhao, L. S. Wang, W. Xia, and Q. W. Yin contributed equally to this work.

\noindent $^\dag$ E-mail: hlei$@$ruc.edu.cn\\
\noindent $^\ddag$ E-mail: guoyf$@$shanghaitech.edu.cn\\
\noindent $^\sharp$ E-mail: yangxiaofan$@$fudan.edu.cn\\
\noindent $^\S$ E-mail: yinjx$@$sustech.edu.cn\\
\noindent $^*$ E-mail: shiyan$\_$li$@$fudan.edu.cn

\end{document}